\documentclass{article}
\usepackage{fuzz}
\usepackage{url}
\usepackage{listings}
\usepackage{changebar}
\usepackage{graphicx}

\title{Recursion in RDF Data Shape Languages}
\author{Arthur Ryman, {\tt arthur.ryman@gmail.com}}
\date{\today}

\begin{document}
\bibliographystyle{acm}
\nochangebars

\maketitle

\begin{abstract}
An RDF data shape is a description of the expected contents of an RDF document (aka graph) or dataset.
A major part of this description is the set of constraints that the document or dataset is required to satisfy.
W3C recently (2014) chartered the RDF Data Shapes Working Group to define SHACL, a standard RDF data shape language.
We refer to the ability to name and reference shape language elements as recursion.
This article provides a precise definition of the meaning of recursion as used in Resource Shape 2.0.
The definition of recursion presented in this article is largely independent of language-specific details.
We speculate that it also applies to ShEx and to all three of the current proposals for SHACL.
In particular, recursion is not permitted in the SHACL-SPARQL proposal, but we conjecture
that recursion could be added by using the definition proposed here as a top-level control structure.
\end{abstract}

\section{Introduction}
\label{sec-intro}
An RDF {\em data shape} is a description of the expected contents of an RDF document (aka graph) or dataset.
A major part of this description is the set of constraints that the document or dataset is required to satisfy.
In this respect, data shapes do for RDF what XML Schema\cite{w3c:xsd11} does for XML.
The term {\em shape} is used instead of {\em schema} to avoid confusion with RDF Schema\cite{w3c:rdfs11} which, like OWL\cite{w3c:owl2},
describes inference rules, not constraints.

W3C recently (2014) chartered the RDF Data Shapes Working Group to define SHACL, a standard RDF data shape language\cite{w3c:shapeswg}.
Both of the member submissions to this working group, Resource Shape 2.0\cite{arthur:rs} and 
Shape Expressions (ShEx) \cite{harold:shex-def} allow shapes to refer to each other.
For example, in Resource Shape 2.0 the property {\tt oslc:valueShape} lets one resource shape refer to another.
ShEx has a similar feature. In these languages, a shape may refer directly or indirectly to itself.

We refer to the ability to name and reference shape language elements as {\em recursion} in analogy with that ubiquitous feature of programming languages which allows a function to call other functions, including itself.
Of course, when writing a recursive function care must be taken to ensure that recursion terminates.
Similarly, when defining a shape language care must be taken to spell out the precise meaning of recursion.
Neither of the member submissions included a precise definition of recursion.

This article provides a precise definition of the meaning of recursion as used in Resource Shape 2.0.
Precision is achieved through the use of Z Notation \cite{spivey:zrm}, a formal specification language based on
typed set theory.
The \LaTeX\ source for this article has been type-checked using the \fuzz\ type-checker \cite{spivey:fuzz}
and is available in the GitHub repository {\tt agryman:shape-recursion} \cite{agryman:shape-recursion}.

The definition of recursion presented in this article is largely independent of language-specific details.
We speculate that it also applies to ShEx and to all three of the current proposals for SHACL.
In particular, recursion is not permitted in the SHACL-SPARQL proposal \cite{peter:shacl}, but we conjecture
that recursion could be added by using the definition proposed here as a top-level control structure.

\subsection{Organization of this Article}
The remainder of this article is organized as follows.
\begin{itemize}
\item Section~\ref{sec-examples} introduces examples in order to ground the following definitions.
\item Section~\ref{sec-basics} defines a few basic RDF concepts.
\item Section~\ref{sec-neighbours} defines neighbour functions and graphs which form the basis for the following definition of recursion.
\item Section~\ref{sec-constraints} defines constraints.
\item Section~\ref{sec-shapes} defines recursive shapes.
\item Section~\ref{sec-languages} discusses how the proposed definition of recursion relates to the existing and proposed shape languages.
\item Section~\ref{sec-conclusion} concludes the article.
\end{itemize}

\section{Examples}
\label{sec-examples}

This section introduces two examples of recursive shapes.

The first recursive shape describes the data in a Personal Information Management application.
This application is highly simplified and easy to understand. 
It is used as a running example to illustrate the formal definitions.
Although this shape is written using recursion, it can be re-written as an equivalent, non-recursive shape.

The second recursive shape describes what it means to be a Polentoni \cite{peter:polentoni}.
This shape is also highly simplified but cannot be re-written as non-recursive using the Resource Shape 2.0 specification.

\subsection{Example: Personal Information Management}
\label{sec-pim}
We use a highly simplified running example to illustrate the concepts defined in the following sections.
Each formal definition is instantiated with data drawn from the running example in order to help the reader understand the
formalism and relate it to RDF.
Although the inclusion of examples lengthens the presentation, we hope that it will make the formalism more tangible and accessible
to readers who are unfamiliar with Z Notation.

Consider a Linked Data \cite{tbl:ld} application for Personal Information Management (PIM).
The application manages documents that contain information about a {\em contact} person and their {\em associates}.
\cbstart
As a Linked Data application, the PIM application provides a REST API for creating, retrieving, updating, and deleting contact information
over HTTP using RDF representations of the data.
Shapes are useful in this context for two main reasons.
First, the PIM application may publish shapes that describe the contact information so that application developers who want to use
the REST API understand the API contract.
Second, the PIM application may internally use a shape engine that automatically validates the data, especially incoming creation and update requests.

The prefixes {\tt rdf:} and {\tt foaf:} as used for terms in the RDF\cite{w3c:rdf11} and FOAF\cite{foaf:spec} vocabularies.
The application maintains the following integrity constraints.
\cbend
\begin{itemize}
\item Each document contains information about exactly one contact person and zero or more of their associates.
\cbstart
A contact person is never an associate of themself.
\cbend
\item Each contact has type {\tt foaf:Person} and has exactly one name given by the property {\tt foaf:name}.
\item The contact's associates are given by the property {\tt foaf:knows} which may have zero or more values.
\item Each associate has type {\tt foaf:Person} and has exactly one name given by {\tt foaf:name}.
\item Each associate is known by exactly one contact given by following the property {\tt foaf:knows} in the backward direction, i.e. the associate is the object of the property and the contact is the subject.
\end{itemize}

Note that these constraints are circular since the definition of contact refers to the definition of associate, and conversely.
We have an obligation to give this circularity a precise meaning.

These constraints are illustrated by a valid document for Alice (Listing~\ref{alice}) and an invalid document for Bob (Listing~\ref{bob}).
All RDF source code examples are written in Turtle format \cite{w3c:turtle11}.

The following document about Alice satisfies all the constraints of the application.
\begin{lstlisting}[caption={Contact document for Alice},label=alice]
# http://example.org/contacts/alice
@prefix foaf: <http://xmlns.com/foaf/0.1/> .
@base <http://example.org/contacts/> .

<alice#me> a foaf:Person ;
	foaf:name "Alice" ;
	foaf:knows
		<bob#me> ,
		<charlie#me> .

<bob#me> a foaf:Person ;
	foaf:name "Bob" .

<charlie#me> a foaf:Person ;
	foaf:name "Charlie" .
\end{lstlisting}

The following document about Bob violates some of the constraints of the application.
\begin{lstlisting}[caption={Contact document for Bob},label=bob]
# http://example.org/contacts/bob
@prefix foaf: <http://xmlns.com/foaf/0.1/> .
@base <http://example.org/contacts/> .

<bob#me> a foaf:Person ;
	foaf:name "Bob" ;
	foaf:knows
		<alice#me> ,
		<charlie#me> .

<alice#me> foaf:name "Alice" .

<charlie#me> a foaf:Person .
\end{lstlisting}

\cbstart
It is clear that the document about Bob is invalid, since Alice has no type and Charlie has no name.

It is also intuitively clear that the document about Alice is valid.
However, if we naively translate the PIM constraints into logical conditions on the document about Alice, then we run into a problem.
All the constraints about types, names, and who knows who are satisfied and unproblematic, but the
constraints about what it means to be a contact or an associate are circular.
A naive translation of these constraints on the Alice document is as follows.
\begin{itemize}
\item If Bob is an associate and Charlie is an associate then Alice is a contact.
\item If Alice is a contact then Bob is an associate.
\item If Alice is a contact then Charlie is an associate.
\end{itemize}

Table~\ref{vars-meaning} introduces propositional variables to stand for statements about being a contact or associate
in the Alice document.
\begin{table}[h]
\begin{center}
\begin{tabular}{|c|c|}
\hline
Variable			& Meaning \\
\hline
$A$	& Alice is a contact. \\
$B$	& Bob is an associate. \\
$C$	& Charlie is an associate. \\
\hline
\end{tabular}
\end{center}
\caption{Meaning of propositional variables in the Alice document}
\label{vars-meaning}
\end{table}

The PIM constraints on the Alice document translate to the following consistency condition.
\[
(B \land C \implies A) \land (A \implies B) \land (A \implies C)
\]

Unfortunately, this consistency condition does not uniquely determine the values of the propositional variables.
In fact, this consistency condition has several solutions as shown in Table~\ref{sol-pim-constraints}.
Only the solution in which all the propositional variable are true agrees with our intuition.
\begin{table}[h]
\begin{center}
\begin{tabular}{|c|c|c|}
\hline
$A$	& $B$	& $C$  \\
\hline
true	& true	& true \\
false	& true	& false \\
false	& false	& true \\
false	& false	& false \\
\hline
\end{tabular}
\end{center}
\caption{Solutions to PIM constraints in the Alice document}
\label{sol-pim-constraints}
\end{table}

This analysis shows that the naive translation of the constraints about contacts and associates
produces a necessary, but not sufficient, consistency condition on the meaning of these constraints.
A precise definition for this type of constraint is given in Section~\ref{sec-shapes}. 
A brief overview of this definition follows.

The correct interpretation of the constraints is based on the observation that they specify two essentially different kinds of information. 
One kind defines rules for labelling nodes with names.
The other kind defines a set of conditions associated with each name and asserts that these conditions must hold at each node labelled with that name.

In the PIM application, the names are {\em contact} and {\em associate}. 
The rules for labelling the nodes in a document are as follows.
\begin{enumerate}
\item Initially, no node has any labels.
\item Start with the node that corresponds to the person that the document is about, and label it as a contact.
\item For each node labelled as a contact, find all the nodes they know, and add an associate label to each of them.
\item For each node labelled as an associate, find all the nodes that they are known by and add a contact label each of them.
\item Repeat the previous two steps until no new labels are added.
\item Note that this procedure always terminates because the number of nodes is finite and the number of names is finite (2 in this case).
\end{enumerate}

In the Alice document, the labelling procedure results in the nodes being labelled as follows.
\begin{itemize}
\item Alice is labelled as a contact because the document is about Alice and Alice is known by Bob and Charlie.
\item Bob is labelled as an associate because Alice knows Bob.
\item Charlie is labelled as an associate because Alice knows Charlie.
\end{itemize}

Whenever a node gets labelled with a name, the conditions associated with the name must hold.
No recursion is involved in this step.

The conditions that must hold for nodes labelled with contact are as follows.
\begin{itemize}
\item A contact must be a person.
\item A contact must have exactly one name.
\item A contact must not know itself.
\end{itemize}

The conditions that must hold for nodes labelled with associate are as follows.
\begin{itemize}
\item An associate must be a person.
\item An associate must have exactly one name.
\item An associate must be known by exactly one node.
\end{itemize}

Although the statement of the PIM constraints uses recursion, the properties of the data in this case allow us to
write an equivalent non-recursive statement \cite{peter:re-recursion}.
Specifically, since a node is an associate only if it is known by a contact, and an associate must be known by exactly one
contact, nothing more is gained by requiring that all nodes that know an associate must be contacts.
Dropping this condition removes the recursion.
However, in general we cannot convert a recursive constraint into an equivalent non-recursive constraint.
The next example illustrates an essentially recursive constraint.

\subsection{Example: Polentoni}
\label{sec-polentoni}
Consider the following definition of what it means to be a Polentoni \cite{peter:polentoni}.
\begin{itemize}
\item A Polentoni lives in exactly one place and that place is Northern Italy.
\item A Polentoni only knows other Polentoni.
\end{itemize}

The definition of Polentoni refers to itself and is therefore recursive.
However, we can give it a precise meaning using the labelling procedure described above.

In this example, the only label name is Polentoni. The labelling procedure is as follows.
\begin{enumerate}
\item Initially, no node has any labels.
\item Start with the node to be checked for being a Polentoni, and label it as a Polentoni.
\item For each node labelled as a Polentoni, find all the nodes they know, and add a Polentoni label to each of them.
\item Repeat the previous step until no new labels are added.
\item Note that this procedure always terminates because the number of nodes is finite and the number of names is finite (1 in this case).
\end{enumerate}

The condition that must hold for nodes labelled with Polentoni is as follows.
\begin{itemize}
\item A Polentoni must live in Northern Italy.
\end{itemize}

Listing~\ref{polentoni-data} contains some sample data.
\begin{lstlisting}[caption={Polentoni sample data},label=polentoni-data]
@prefix ex: <http://example.org/polentoni#> .

ex:Enrico ex:livesIn ex:NorthernItaly .
ex:Diego ex:livesIn  ex:NorthernItaly .
ex:Alessandro ex:livesIn ex:NorthernItaly .
ex:Sergio ex:livesIn ex:NorthernItaly .
ex:John ex:livesIn ex:NorthernItaly .
ex:Maurizio ex:livesIn ex:SouthernItaly .

ex:Enrico ex:knows ex:John .
ex:John ex:knows ex:Maurizio .
ex:Diego ex:knows ex:Alessandro .
ex:Alessandro ex:knows ex:Diego .
ex:Alessandro ex:knows ex:Sergio .
\end{lstlisting}

\begin{figure}[h]
\centering
\includegraphics[scale=0.5]{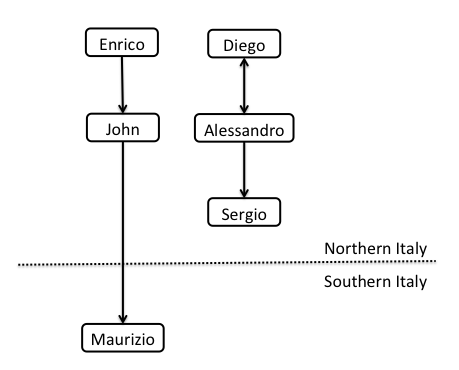}
\caption{Polenti sample data}
\label{fig:polentoni-data}
\end{figure}

Figure~\ref{fig:polentoni-data} depicts the Polentoni sample data where, for example, the arrow from Enrico to John
indicates that Enrico knows John.

Checking Enrico results in Enrico, John, and Maurizio being labelled as Polentoni. 
However, Maurizio lives in Southern Italy so Enrico is not a Polentoni.

Checking Diego results in Diego, Alessandro, and Sergio begin labelled as Polentoni.
They all live in Northern Italy so Diego is a Polentoni.

Note that if Resource Shape 2.0 were more expressive then we could rewrite the definition of Polentoni to avoid recursion as follows.
\begin{itemize}
\item A Polentoni lives in Northern Italy (and nowhere else).
\item Everyone that a Polentoni knows, directly or indirectly, lives in Northern Italy (and nowhere else).
\end{itemize}
The price paid for eliminating recursion is that now we have introduced the transitive closure of the {\em knows} relation, which is beyond the expressive power of the Resource Shape 2.0 specification.

Transitive closure is, however, expressible using SPARQL property paths.
In fact, all the Polentoni constraints can be expressed by a single SPARQL query.
Listing~\ref{polentoni-rq} contains a SPARQL query that finds all non-Polentoni people in a graph,
where we assume that a person is any resource that lives somewhere, or knows someone, or is known by someone.
Note the use of the property path {\tt ex:knows*} which is referred to as a {\tt ZeroOrMorePath} expression.
\begin{lstlisting}[caption={SPARQL query for non-Polentoni people},label=polentoni-rq]
# polentoni.rq

prefix ex: <http://example.org/polentoni#>

# finds all non-Polentoni person nodes ?this in the graph
select distinct ?this
where {
	# binds each person node to ?this
	{
		select distinct ?this
		where {
			{?this ex:livesIn ?region}
			union
			{?this ex:knows ?person}
			union
			{?person ex:knows ?this}
		}
	}

	# binds each person that ?this knows, 
	# directly or indirectly, to ?person
	?this ex:knows* ?person .

	# A non-Polentoni ?person must not live 
	# in and only in Northern Italy
	{
		# ?person lives nowhere
		filter not exists {?person ex:livesIn ?region}
	}
	union
	{
		# ?person lives somewhere not Northern Italy
		?person ex:livesIn ?region.
		filter (?region != ex:NorthernItaly)
	}
}
\end{lstlisting}

Table~\ref{polentoni-results} gives the results of running the non-Polenoni query on the data contained in Listing~\ref{polentoni-data}.
\begin{table}[h]
\begin{center}
\begin{tt}
\begin{tabular}{|c|}
\hline
this \\
\hline
http://example.org/polentoni\#Maurizio \\
http://example.org/polentoni\#John \\
http://example.org/polentoni\#Enrico \\
\hline
\end{tabular}
\end{tt}
\end{center}
\caption{SPARQL query results for non-Polentoni people}
\label{polentoni-results}
\end{table}
\begin{itemize}
\item Maurizio is non-Polentoni because he lives in Southern Italy.
\item John is non-Polentoni because he knows Maurizio.
\item Enrico is non-Polentoni because he knows John.
\end{itemize}

One might therefore contemplate avoiding the issue of recursion by adding powerful path expressions to the shape language.
However, it is unclear that path expressions alone are sufficiently powerful to cover all the cases currently expressible in Resource Shape 2.0.
Furthermore, even if that were true, translating recursive references into property path expressions would impose a severe burden on the shape author.
The use of recursion allows concise and intuitively clear descriptions so, as long as recursion can be given a precise definition, 
there is good reason to include in future shape languages.

\cbend

\section{Basic RDF Concepts}
\label{sec-basics}

This section formalizes some basic RDF concepts.
For full definitions consult the RDF specification\cite{w3c:rdf11}.

\subsection{Terms}

Let $TERM$ be the set of all RDF {\em terms}.
\begin{zed}
[TERM]
\end{zed}

The set of all RDF terms is partitioned into {\em IRIs}, {\em blank nodes}, and {\em literals}.
\begin{axdef}
IRI, BNode, Literal: \power TERM
\where
\langle IRI, BNode, Literal \rangle \partition TERM
\end{axdef}

For example, the documents for Alice and Bob contain the following distinct literals
where $Alice$ denotes {\tt "Alice"}, etc.
\begin{axdef}
	Alice, Bob, Charlie: Literal
\where
	\disjoint \langle \{Alice\}, \{Bob\}, \{Charlie\} \rangle
\end{axdef}
and the following distinct IRIs where $alice$ denotes {\tt http://example.org/contacts/alice\#me}, etc., 
$rdf\_type$ denotes {\tt rdf:type}, and
$foaf\_Person$ denotes {\tt foaf:Person}, etc.
\begin{axdef}
	alice, bob, charlie: IRI \\
	rdf\_type: IRI \\
	foaf\_Person, foaf\_name, foaf\_knows: IRI
\where
	\disjoint \langle \{alice\}, \{bob\}, \{charlie\}, \{rdf\_type\}, \\
\t1		\{foaf\_Person\}, \{foaf\_name\}, \{foaf\_knows\} \rangle
\end{axdef}

\subsection{Triples}

An RDF {\em triple} is a statement that consists of three terms referred to as {\em subject}, {\em predicate}, and {\em object}.
\begin{zed}
Triple == \{~ s, p, o: TERM | s \notin Literal \land p \in IRI ~\}
\end{zed}
\begin{itemize}
\item The subject must not be a literal.
\item The predicate must be an IRI.
\end{itemize}

For example, the statement that Alice is a person is represented by the following triple.
\[\vdash 
	(alice, rdf\_type, foaf\_Person) \in Triple
\]

\subsection{Graphs}

It is common to visualize a triple as a directed arc from the subject to the object, labelled by the predicate.
A set of triples may therefore may visualized as a directed graph (technically, a directed, labelled, multigraph).
We are only concerned with finite graphs here.

An RDF {\em graph} is a finite set of triples.
\begin{zed}
Graph == \finset Triple
\end{zed}

For example, the following graph contains the triples in the document about Alice.
\begin{axdef}
	alice\_graph: Graph
\where
	alice\_graph = \\
\t1		\{ (alice, rdf\_type, foaf\_Person), \\
\t1		(alice, foaf\_name, Alice), \\
\t1		(alice, foaf\_knows, bob), \\
\t1		(alice, foaf\_knows, charlie), \\
\t1		(bob, rdf\_type, foaf\_Person), \\
\t1		(bob, foaf\_name, Bob), \\
\t1		(charlie, rdf\_type, foaf\_Person), \\
\t1		(charlie, foaf\_name, Charlie) \}
\end{axdef}

\begin{figure}[h]
\centering
\includegraphics[scale=0.5]{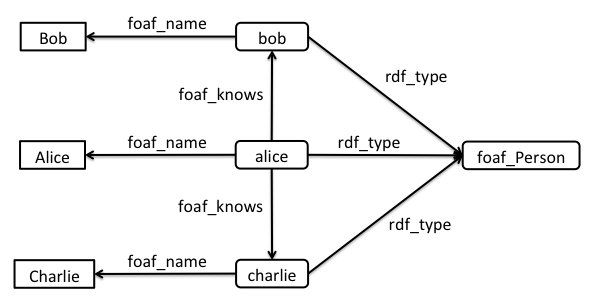}
\caption{Alice contact graph}
\label{fig:alice-contact}
\end{figure}

Figure~\ref{fig:alice-contact} depicts the document about Alice as a directed, labelled graph.

It is convenient to define functions that map graphs to the sets of subjects, predicates, and objects that appear in the graph.
\begin{zed}
subjects == (\lambda g: Graph @ \{~ s, p, o: TERM | (s,p,o) \in g @ s ~\})
\also
predicates == (\lambda g: Graph @ \{~ s, p, o: TERM | (s,p,o) \in g @ p ~\})
\also
objects == (\lambda g: Graph @ \{~ s, p, o: TERM | (s,p,o) \in g @ o ~\}) 
\end{zed}

For example, the graph for Alice contains the following predicates.
\[\vdash 
	predicates(alice\_graph) = \\
\t1		\{ rdf\_type, foaf\_name, foaf\_knows \}
\]

The {\em nodes} of a graph are its subjects and objects.
\begin{zed}
nodes == (\lambda g: Graph @ subjects(g) \cup objects(g))
\end{zed}

For example, the graph for Alice contains the following nodes.
\[\vdash 
	nodes(alice\_graph) = \\
\t1		\{ alice, bob, charlie, Alice, Bob, Charlie, foaf\_Person \}
\]

A {\em pointed graph} consists of a graph and a {\em base node} in the graph.
\begin{schema}{PointedGraph}
	graph: Graph \\
	baseNode: TERM
\where
	baseNode \in nodes(graph)
\end{schema}
\begin{itemize}
\item The base node is some node in the graph.
\end{itemize}

The base node of a pointed graph is also referred to as the {\em start node} or {\em focus node} of the graph, depending on the context.

For example, $alice$ is the natural base node of the graph for Alice.
\begin{axdef}
	alice\_pg: PointedGraph
\where
	alice\_pg.graph = alice\_graph
\also
	alice\_pg.baseNode = alice
\end{axdef}
\begin{itemize}
\item The graph is $alice\_graph$.
\item The base node is $alice$.
\end{itemize}

\section{Neighbour Functions}
\label{sec-neighbours}

RDF applications often impose conditions on nodes, and related conditions on their {\em neighbours},
where a neighbour is some node that bears a specified relation to the given node.
When the neighbour relation between nodes is specified by traversing triples, we say that the nodes
are connected by a {\em path}. 
SPARQL 1.1\cite{w3c:sparql11} defines a {\em property path} syntax for specifying paths. 

More generally, applications may use neighbour relations that cannot be specified by property paths.
Many such relations might be specified by SPARQL queries that bind pairs of variables to nodes.
For maximum generality, we do not place restrictions on how neighbour relations are specified.

A {\em neighbour function} is any mapping from graphs to pair of nodes that belong to the graph.
\begin{axdef}
	Neighbour: \power (Graph \fun (TERM \rel TERM))
\where
	Neighbour = \\
\t1		\{~ q: Graph \fun (TERM \rel TERM) | \\
\t2			(\forall g: Graph @ q(g) \subseteq \{~ x, y: nodes(g) ~\}) ~\}
\end{axdef}
\begin{itemize}
\item A neighbour function is a mapping that maps a graph $g$ to a binary relation on the nodes of $g$.
\end{itemize}

We say that the pair of nodes $(x,y)$ {\em matches} the neighbour function $q$ in the graph $g$ when $(x,y) \in q(g)$.

\subsection{Simple Path Expressions}

Simple path expressions define a very commonly used type of neighbour function.

A predicate $p$ defines a simple path expression $forward(p)$ by traversing triples in the forward direction.
Forward path expressions are referred to as {\tt PredicatePath} expressions in SPARQL 1.1.
\begin{axdef}
	forward: IRI \fun Neighbour
\where
	\forall p: IRI; g: Graph @ \\
\t1		forward(p)(g) = \\
\t2			\{~ s, o: nodes(g) | (s,p,o) \in g ~\}
\end{axdef}
\begin{itemize}
\item The simple path expression $forward(p)$ matches all pairs $(s,o)$ such that $(s,p,o)$ is a triple in $g$.
\end{itemize}

For example, the following are forward path expressions.
\begin{zed}
	has\_type == forward(rdf\_type)
\also
	has\_name == forward(foaf\_name)
\also
	knows == forward(foaf\_knows)
\end{zed}

The forward path expression $has\_type$ matches the following pairs of nodes in the graph for Alice.
\[\vdash
	has\_type(alice\_graph) = \\
\t1		\{ (alice, foaf\_Person), \\
\t1		(bob, foaf\_Person), \\
\t1		(charlie, foaf\_Person) \}
\]

Similarly, a predicate $p$ defines a simple path expression $backward(p)$ by traversing triples in the backward direction.
Backward path expressions are referred to as {\tt InversePath} expressions in SPARQL 1.1.
\begin{axdef}
	backward: IRI \fun Neighbour
\where
	\forall p: IRI; g: Graph @ \\
\t1		backward(p)(g) = \\
\t2			\{~ o, s: nodes(g) | (s,p,o) \in g ~\}
\end{axdef}
\begin{itemize}
\item The simple path expression $backward(p)$ matches all pairs $(o,s)$ such that $(s,p,o)$ is a triple in $g$.
\end{itemize}

For example, the following is a backward path expression.
\begin{zed}
	is\_known\_by == backward(foaf\_knows)
\end{zed}

The backward path expression $is\_known\_by$ matches the following pairs of nodes in the graph for Alice.
\[\vdash 
	is\_known\_by(alice\_graph) = \\
\t1		\{ (bob, alice), \\
\t1		(charlie, alice) \}
\]

\subsection{Values}

Given a graph $g$ and a node $x \in nodes(g)$, the set of all nodes that can be reached from $x$ by matching the neighbour function $q$  is $values(g,x,q)$.
\begin{axdef}
values: Graph \cross TERM \cross Neighbour \fun \finset TERM
\where
\forall g: Graph; x: TERM; q: Neighbour @ \\
\t1	values(g,x,q) = \{~ y: nodes(g) | (x,y) \in q(g) ~\}
\end{axdef}
\begin{itemize}
\item The node $y$ is in $values(g,x,q)$ when $(x,y)$ matches $q$ in $g$.
\end{itemize}

For example, in the graph for Alice the node $alice$ and forward path expression $knows$ have the following values.
\[\vdash 
	values(alice\_graph, alice, knows) = \{bob, charlie\}
\]

\section{Constraints}
\label{sec-constraints}

RDF applications often impose constraints on the data graphs they process.
A given graph either {\em satisfies} or {\em violates} the constraint.
Thus a constraint partitions the set of all graphs into two disjoint subsets,
namely the set of all graphs that satisfy the constraint and the set of all graphs that violate the constraint.
A constraint is therefore defined by the set of graphs that satisfy it.

A {\em constraint} is a, possibly infinite, set of graphs. 
\begin{zed}
	Constraint == \power Graph
\end{zed}

For example, suppose we define a {\em small graph} to be a graph that has at most 10 triples.
The set of all small graphs is a constraint.
\begin{axdef}
	small\_graphs: Constraint
\where
	small\_graphs = \{~ g: Graph | \# g \leq 10 ~\}
\end{axdef}

The Alice graph satisfies this constraint.
\[\vdash
	alice\_graph \in small\_graphs
\]

\subsection{Node Constraints}

A {\em parameterized constraint} is a mapping from 
some parameter set $X$ to constraints.
\begin{zed}
	ParameterizedConstraint[X] == X \fun Constraint
\end{zed}

A {\em term constraint} is a constraint that is parameterized by terms.
\begin{zed}
	TermConstraint == ParameterizedConstraint[TERM]
\end{zed}

For example, given a term $x \in TERM$, the constraint $hasSubject(x)$ is the set of all graphs that have $x$ as a subject.
\begin{axdef}
	hasSubject: TermConstraint
\where
\forall x: TERM @ \\
\t1	hasSubject(x) = \{~ g: Graph | x \in subjects(g) ~\}
\end{axdef}

Similarly, $hasPredicate(x)$, $hasObject(x)$, and $hasNode(x)$ are constraints with the analogous definitions.
\begin{zed}
	hasPredicate == (\lambda x: TERM @ \{~ g: Graph | x \in predicates(g) ~\})
\also
	hasObject == (\lambda x: TERM @ \{~ g: Graph | x \in objects(g) ~\})
\also
	hasNode == (\lambda x: TERM @ \{~ g: Graph | x \in nodes(g) ~\})
\end{zed}

Note that $hasNode(x)$ is the union of $hasSubject(x)$ and $hasObject(x)$.
\[\vdash 
	\forall x: TERM @ \\
\t1		hasNode(x) = hasSubject(x) \cup hasObject(x)
\]

A {\em node constraint} is a term constraint in which the term is a node in each graph that satisfies the constraint.
\begin{axdef}
	NodeConstraint: \power TermConstraint
\where
	NodeConstraint = \\
\t1 \{~ c: TermConstraint | \forall x: TERM @ \forall g: c(x) @ x \in nodes(g) ~\}
\end{axdef}

For example, $hasNode$ is a node constraint.
\[\vdash
	hasNode \in NodeConstraint
\]

The PIM application enforces the following node constraints.

Both contact and associate nodes must be people.
\begin{axdef}
	is\_a\_person: NodeConstraint
\where
	\forall x: TERM @ \\
\t1		is\_a\_person(x) = \\
\t2		\{~ g: Graph | (x, rdf\_type, foaf\_Person) \in g ~\}
\end{axdef}
\begin{itemize}
\item A node {\em is a person} when it has a {\tt foaf:Person} as one of its RDF types.
\end{itemize}

For example, the Alice graph satisfies this constraint at the $alice$, $bob$, and $charlie$ nodes.
\[\vdash
	alice\_graph \in is\_a\_person(alice) \land \\
	alice\_graph \in is\_a\_person(bob) \land \\
	alice\_graph \in is\_a\_person(charlie)
\]

\cbstart
Both contact and associate nodes must have exactly one name.
\cbend
\begin{axdef}
	has\_one\_name: NodeConstraint
\where
	\forall x: TERM @ \\
\t1		has\_one\_name(x) = \\
\t2			\{~ g: Graph | \exists_1 y: TERM @ (x, foaf\_name, y) \in g ~\}
\end{axdef}
\begin{itemize}
\item A node {\em has one name} when it is the subject of exactly one {\tt foaf:name} triple.
\end{itemize}

For example, the Alice graph satisfies this constraint at the $alice$, $bob$, and $charlie$ nodes.
\[\vdash
	alice\_graph \in has\_one\_name(alice) \land \\
	alice\_graph \in has\_one\_name(bob) \land \\
	alice\_graph \in has\_one\_name(charlie)
\]

\cbstart
Associate nodes must be known by exactly one node.
\cbend
\begin{axdef}
	is\_known\_by\_one: NodeConstraint
\where
	\forall x: TERM @ \\
\t1		is\_known\_by\_one(x) = \\
\t2			\{~ g: Graph | \exists_1 y: TERM @ (y, foaf\_knows, x) \in g ~\}
\end{axdef}
\begin{itemize}
\item A node {\em is known by one} node when it is the object of exactly one {\tt foaf:knows} triple.
\end{itemize}

For example, the Alice graph satisfies this constraint at the $bob$ and $charlie$ nodes.
\[\vdash
	alice\_graph \in is\_known\_by\_one(bob) \land \\
	alice\_graph \in is\_known\_by\_one(charlie)
\]

A contact node must satisfy the following constraint.
\begin{axdef}
	contact\_nc: NodeConstraint
\where
	\forall x: TERM @ \\
\t1		contact\_nc(x) = \\
\t2			is\_a\_person(x) \cap \\
\t2			has\_one\_name(x)
\end{axdef}
\begin{itemize}
\item A contact is a person and has one name.
\end{itemize}

The Alice graph satisfies this constraint at the $alice$, $bob$, and $charlie$ nodes.
\[\vdash
	alice\_graph \in contact\_nc(alice) \land \\
	alice\_graph \in contact\_nc(bob) \land \\
	alice\_graph \in contact\_nc(charlie)
\]

An associate node must satisfy the following constraint.
\begin{axdef}
	associate\_nc: NodeConstraint
\where
	\forall x: TERM @ \\
\t1		associate\_nc(x) = \\
\t2			is\_a\_person(x) \cap \\
\t2			has\_one\_name(x) \cap \\
\t2			is\_known\_by\_one(x)
\end{axdef}
\begin{itemize}
\item An associate is a person, has one name, and is known by one node.
\end{itemize}

The Alice graph satisfies this constraint at the $bob$ and $charlie$ nodes.
\[\vdash
	alice\_graph \in associate\_nc(bob) \land \\
	alice\_graph \in associate\_nc(charlie)
\]

\section{Shapes}
\label{sec-shapes}

In general, a shape is any description of the expected contents of a graph.
In this article we deal only with shapes that describe graphs using the following structure.
A shape is a structure that defines how to associate a set of node constraints with each node of a data graph in two steps.
\begin{enumerate}
\item Label each node of the graph with a set of node constraint names using a set of neighbour functions.
\item Map each name to a node constraint.
\end{enumerate}
These steps are described in detail below.

Note that this definition of shape is very prescriptive about the labelling process but is completely independent of the details of 
both the neighbour functions and the node constraints.
We speculate that the labelling process can be used to handle the recursive aspects of a wide variety of shape languages that differ only in their 
expressiveness for defining neighbour functions and node constraints.
For example, Resource Shape 2.0 uses forward and backward path expressions as neighbour functions and has a small, fixed set of simple node constraints. 
SHACL-SPARQL allows node constraints to be expressed by arbitrary SPARQL 1.1 queries, but does not
allow explicit recursion.

\subsection{Labelling Data Graph Nodes with Constraint Names}
\cbstart
A shape contains a set of named constraints.
A constraint may refer to other constraints by name.
This means that a constraint may refer directly or indirectly to itself, in which case the constraint is recursive.
\cbend

Shapes themselves may be represented as RDF graphs, so it is tempting to use IRIs to name node constraints.
However, we introduce a new given set of names to emphasize that this set is logically independent of how we represent shapes.
\begin{zed}
	[NAME]
\end{zed}

For example, there are two distinct kinds of node in the PIM application, namely contact and associate.
\begin{axdef}
	contact, associate: NAME
\where
	contact \neq associate
\end{axdef}
\begin{itemize}
\item $contact$ and $associate$ are distinct names.
\end{itemize}

Since graphs appear in several roles, there is scope for confusion.
To clarify its role, the graph to which constraints are being applied will be referred to as the {\em data graph}.

The part of a shape that defines how data graph nodes are labelled is a {\em neighbour graph}.
A neighbour graph is a directed, labelled, multigraph whose nodes are names and whose
arcs are labelled by neighbour functions.
\begin{schema}{NeighbourGraph}
	names: \finset NAME \\
	arcs: \finset (NAME \cross Neighbour \cross NAME)
\where
	arcs \subseteq names \cross Neighbour \cross names
\end{schema}
\begin{itemize}
\item The nodes are names and the arcs are labelled by neighbour functions.
\end{itemize}

For example, in the PIM application, contacts are related to associates by the $knows$ forward path expression, and associates
are related to contacts by the $is\_known\_by$ backward path expression.
\begin{axdef}
	pim\_ng : NeighbourGraph
\where
	pim\_ng.names = \{ contact, associate \}
\also
	pim\_ng.arcs = \\
\t1		\{ (contact, knows, associate), \\
\t1		(associate, is\_known\_by, contact) \}
\end{axdef}

A {\em pointed neighbour graph} consists of a neighbour graph and a {\em base name} in the graph.
\begin{schema}{PointedNeighbourGraph}
	NeighbourGraph \\
	baseName: NAME
\where
	baseName \in names
\end{schema}
\begin{itemize}
\item The base name belongs to the graph.
\end{itemize}

The base name of a pointed neighbour graph is also referred to as the {\em start name} or {\em focus name}, depending on the context.

For example, $contact$ is the natural base name in the PIM application.
\begin{axdef}
	pim\_png: PointedNeighbourGraph
\where
	pim\_png.names = pim\_ng.names
\also
	pim\_png.arcs = pim\_ng.arcs
\also
	pim\_png.baseName = contact
\end{axdef}

A {\em named node} is pair of the form $(x,a)$ where $x$ is a data graph node and $a$ is a node constraint name.
\begin{zed}
NamedNode == TERM \cross NAME
\end{zed}

For example, $(alice, contact)$ is named node.
\[\vdash
	(alice,contact) \in NamedNode
\]

A data graph $g$ and a neighbour graph $ng$ define a {\em requires} binary relation $requires(g,ng)$ on the set of named nodes.
The meaning of this relation is that if $(x,a) \inrel{requires} (y,b)$ then whenever $x$ must satisfy the constraints named by $a$ then
$y$ must satisfy the constraints named by $b$.
\begin{axdef}
requires: Graph \cross NeighbourGraph \fun (NamedNode \rel NamedNode)
\where
\forall g: Graph; ng: NeighbourGraph @ \\
\t1 requires(g,ng) = \\
\t2		\{~ x, y: nodes(g); a, b: NAME; q: Neighbour | \\
\t3			(a,q,b) \in ng.arcs \land \\
\t3			(x,y) \in q(g) @ \\
\t4				(x,a) \mapsto (y,b) ~\}
\end{axdef}
\begin{itemize}
\item The named node $(x,a)$ requires $(y,b)$ when the neighbour graph includes an arc $(a,q,b)$ and the node $y$ can be
reached from $x$ by matching the neighbour function $q$ in $g$.
\end{itemize}

For example, the requires relation for the Alice graph in the PIM application is as follows.
\[\vdash
	requires(alice\_graph, pim\_ng) = \\
\t1		\{ (alice,contact) \mapsto (bob,associate), \\
\t1		(alice,contact) \mapsto (charlie,associate),\\
\t1		(bob,associate) \mapsto (alice,contact), \\
\t1		(charlie,associate) \mapsto (alice,contact) \}
\]

A {\em labelled graph} is a data graph whose nodes are each labelled by a, possibly empty, set of names.
\begin{schema}{LabelledGraph}
graph: Graph \\
names: \finset NAME \\
label: TERM \pfun \finset NAME
\where
label \in nodes(graph) \fun \finset names
\end{schema}
\begin{itemize}
\item Each node in the graph is labelled by a set of names.
\end{itemize}

For example, the following is a labelled graph based on the Alice graph.
\begin{axdef}
	alice\_lg: LabelledGraph
\where
	alice\_lg.graph = alice\_graph
\also
	alice\_lg.names = \{ contact, associate \}
\also
	alice\_lg.label = \\
\t1		\{ alice \mapsto \{ contact \}, \\
\t1		bob \mapsto \{ associate \}, \\
\t1		charlie \mapsto \{ associate \}, \\
\t1		Alice \mapsto \emptyset, \\
\t1		Bob \mapsto \emptyset, \\
\t1		Charlie \mapsto \emptyset, \\
\t1		foaf\_Person \mapsto \emptyset \}
\end{axdef}

A pointed graph and a pointed neighbour graph determine a unique labelled graph.
Intuitively, the labelling process starts by labelling the base node with the base name. 
Next, the neighbour graph is checked for arcs that begin at $baseName$, e.g. $(baseName, q, b)$.
For each such arc compute the set $values(g,q,baseNode)$ and for each node $y$ in this set, label $y$ with $b$.
Now repeat these steps taking $y$ as the new base node and $b$ as the new base name, but only do this
once for each named node $(y,b)$.
Since there are a finite number of nodes and a finite number of names, this process always terminates.
\begin{schema}{LabelGraph}
	PointedGraph \\
	PointedNeighbourGraph \\
	LabelledGraph
\where
	\LET ng == \theta NeighbourGraph @ \\
\t1		\LET R == (requires(graph,ng))\star @ \\
\t2			label = (\lambda y: nodes(graph) @ \\
\t3				\{~ b: names | (baseNode,baseName) \inrel{R} (y,b) ~\})
\end{schema}
\begin{itemize}
\item The label of a node $y$ is the set of names $b$ such that the named node $(y,b)$ is related to the 
base named node $(baseNode,baseName)$ by $R$ the reflexive-transitive closure of the requires relation $requires(graph,shape)$.
\item Note that this labelling process makes use of $R$, the reflexive-transitive closure of the requires relation.
The use of $R$ avoids difficulties associated with explicitly recursive definitions. 
We have, in effect, eliminated explicit recursion by computing a transitive closure of a finite binary relation.
\item Note that the components of $LabelledGraph$ are uniquely determined by the components of $PointedGraph$
and $PointedNeighbourGraph$. 
\end{itemize}

For example, the pointed graph $alice\_pg$ and the pointed neighbour graph $pim\_png$ uniquely determine
the labelled graph $alice\_lg$.
\begin{zed}
	\forall LabelGraph | \\
\t1		\theta PointedGraph = alice\_pg \land \\
\t1		\theta PointedNeighbourGraph = pim\_png @ \\
\t2			\theta LabelledGraph = alice\_lg
\end{zed}

\subsection{Mapping Constraint Names to Node Constraints}

The association of node constraints to graph nodes is given by a mapping.
\begin{schema}{NodeConstraints}
	names: \finset NAME \\
	constraint: NAME \pfun NodeConstraint
\where
	\dom constraint = names
\end{schema}
\begin{itemize}
\item Each name maps to a node constraint.
\end{itemize}

For example, the PIM application associates the following node constraints with names.
\begin{axdef}
	pim\_ncs: NodeConstraints
\where
	pim\_ncs.names = \{ contact, associate \}
\also
	pim\_ncs.constraint = \\
\t1		\{ contact \mapsto contact\_nc, \\
\t1		associate \mapsto associate\_nc \}
\end{axdef}
\begin{itemize}
\item The PIM application has two kinds of nodes, named $contact$ and $associate$.
\item $contact$ nodes must satisfy the $contact\_nc$ node constraint.
\item $associate$ nodes must satisfy the $associate\_nc$ node constraint.
\end{itemize}

A {\em constrained graph} is an assignment of a, possibly empty, set of node constraints to each node of the graph.
\begin{schema}{ConstrainedGraph}
	graph: Graph \\
	constraints: TERM \pfun \finset NodeConstraint
\where
	\dom constraints = nodes(graph)
\end{schema}
\begin{itemize}
\item Each node of the data graph has a set of node constraints.
\end{itemize}

For example, the PIM application enforces the following constraints on the Alice graph.
\begin{axdef}
	alice\_cg: ConstrainedGraph
\where
	alice\_cg.graph = alice\_graph
\also
	alice\_cg.constraints = \\
\t1		\{ alice \mapsto \{ contact\_nc \}, \\
\t1		bob \mapsto \{ associate\_nc \}, \\
\t1		charlie \mapsto \{ associate\_nc \}, \\
\t1		Alice \mapsto \emptyset, \\
\t1		Bob \mapsto \emptyset, \\
\t1		Charlie \mapsto \emptyset, \\
\t1		foaf\_Person \mapsto \emptyset \}
\end{axdef}
\begin{itemize}
\item $alice$ must satisfy the contact node constraint.
\item $bob$ and $charlie$ must satisfy the associate node constraint.
\item There are no node constraints on the remaining nodes.
\end{itemize}

A constrained graph is {\em valid} if it satisfies all the constraints at each node.
\begin{schema}{ValidGraph}
	ConstrainedGraph
\where
	\forall x: nodes(graph) @ \\
\t1		\forall c: constraints(x) @ \\
\t2			graph \in c(x)
\end{schema}
\begin{itemize}
\item A valid data graph satisfies each node constraint at each node.
\end{itemize}

For example, the constrained graph $alice\_cg$ is valid.
\[\vdash
	alice\_cg \in ValidGraph
\]

A mapping from nodes to names ($LabelledGraph$) and a mapping from names to node constraints ($NodeConstraints$) uniquely determines a
mapping from nodes to node constraints ($ConstrainedGraph$).
\begin{schema}{ConstrainGraph}
	LabelledGraph \\
	NodeConstraints \\
	ConstrainedGraph
\where
	\forall x: nodes(graph) @ \\
\t1		constraints(x) = \\
\t2			\{~ a: label(x) @ constraint(a) ~\}
\end{schema}
\begin{itemize}
\item The set of node constraints at each node $x$ of a labelled data graph is equal to the to the set 
node constraints named by the labels $a$ at $x$.
\item Note that the components of $ConstrainedGraph$ are uniquely determined by the components of $LabelledGraph$
and $NodeConstraints$.
\end{itemize}

For example, the Alice labelled graph $alice\_lg$ and the PIM node constraints $pim\_ncs$ uniquely determine the
Alice constrained graph $alice\_cg$.
\[\vdash
	\forall ConstrainGraph | \\
\t1		\theta LabelledGraph = alice\_lg \land \\
\t1		\theta NodeConstraints = pim\_ncs @ \\
\t2			\theta ConstrainedGraph = alice\_cg
\]

\subsection{Shapes as Constraints}

A {\em shape} consists of a neighbour graph and node constraints.
\begin{schema}{Shape}
	NeighbourGraph \\
	NodeConstraints
\end{schema}

For example, the neighbour graph $pim\_ng$ and the node constraints $pim\_nc$ define a shape for the PIM application.
\begin{axdef}
	pim\_shape: Shape
\where
	pim\_shape.names = \{ contact, associate \}
\also
	pim\_shape.arcs = pim\_ng.arcs
\also
	pim\_shape.constraint = pim\_ncs.constraint
\end{axdef}		

A {\em pointed shape} consists of a pointed neighbour graph and node constraints.
\begin{schema}{PointedShape}
	PointedNeighbourGraph \\
	NodeConstraints
\end{schema}

For example, the pointed neighbour graph $pim\_png$, which has base name $contact$, and the node constraints
$pim\_ncs$ define a pointed shape for the PIM application.
\begin{axdef}
	pim\_ps: PointedShape
\where
	pim\_ps.names = \{ contact, associate \}
\also
	pim\_ps.baseName = contact
\also
	pim\_ps.arcs = pim\_ng.arcs
\also
	pim\_ps.constraint = pim\_ncs.constraint
\end{axdef}		

A pointed data graph satisfies a pointed shape if the constrained graph produced by the composition of the labelling 
and constraining processes is valid.
\begin{schema}{SatisfiesShape}
	PointedShape \\
	PointedGraph \\
	LabelGraph \\
	ConstrainGraph
\where
	ValidGraph
\end{schema}
\begin{itemize}
\item The constrained graph that results from the labelling and constraining processes must be valid.
\item Note that the components of $LabelGraph$ and $ConstrainGraph$ are uniquely determined by the components
of $PointedShape$ and $PointedGraph$. The validity condition ($ValidGraph$) therefore determines a relation
between $PointedShape$ and $PointedGraph$.
The pointed graph is said to {\em satisfy} the pointed shape.
\end{itemize}

For example, the pointed Alice graph satisfies the pointed PIM shape.
\[\vdash
	\exists_1 SatisfiesShape @ \\
\t1		\theta PointedShape = pim\_ps \land \\
\t1		\theta PointedGraph = alice\_pg
\]

A pointed shape determines a node constraint.
\begin{axdef}
	shapeConstraint: PointedShape \fun NodeConstraint
\where
	\forall ps: PointedShape; x: TERM @ \\
\t1		shapeConstraint(ps)(x) = \\
\t2			\{~ SatisfiesShape | ps = \theta PointedShape \land x = baseNode @ graph ~\}
\end{axdef}
\begin{itemize}
\item Given a pointed shape $ps$, $shapeConstraint(ps)$ is a node constraint. 
Given a node $x$, $shapeConstraint(ps)(x)$ is the set of all graphs $graph$ such that the pointed graph
formed by using $x$ as the base node satisfies $ps$.
\end{itemize}

For example, the pointed PIM shape defines a node constraint.
\begin{axdef}
	pim\_nc: NodeConstraint
\where
	pim\_nc = shapeConstraint(pim\_ps)
\end{axdef}

The graph $alice\_graph$ satisfies this node constraint at the node $alice$.
\[\vdash
	alice\_graph \in pim\_nc(alice)
\]

\section{Relation to Shape Languages}
\label{sec-languages}

This section discusses how the preceding formalism relates to Resource Shape 2.0, ShEx, and SHACL.

\subsection{Relation to Resource Shape 2.0}

Resource Shape 2.0 provides a small vocabulary for defining simple, commonly occurring node constraints
such as property occurrence, range, and allowed values. These are uncontentious and will not be discussed further.

As mentioned above, Resource Shape 2.0 also allows recursive shapes via the property {\tt oslc:valueShape}.
The preceding formalism was motivated by Resource Shape 2.0 and, not surprisingly, provides a precise
description of the meaning of recursive shapes in that language.

\subsubsection{Example: Personal Information Management}
The following listings illustrate the use of {\tt oslc:valueShape} for the running PIM example.

Listing~\ref{contact} contains the resource shape for contacts.
\cbstart
Note that Resource Shape 2.0 is incapable of expressing the constraint that a contact must not have itself as an associate.
\cbend
\begin{lstlisting}[caption={Resource shape for contacts}, label=contact]
# http://example.org/shapes/contact
@prefix foaf: <http://xmlns.com/foaf/0.1/> .
@prefix oslc: <http://open-services.net/ns/core#> .
@prefix xsd: <http://www.w3.org/2001/XMLSchema#>.
@base <http://example.org/shapes/> .

<contact> a oslc:ResourceShape ;
	oslc:describes foaf:Person ;
	oslc:property 
		<contact#name> , 
		<contact#knows> .

<contact#name> a oslc:Property ;
	oslc:name "name" ;
	oslc:occurs oslc:Exactly-one ;
	oslc:propertyDefinition foaf:name ;
	oslc:valueType xsd:string .

<contact#knows> a oslc:Property ;
	oslc:name "knows" ;
	oslc:occurs oslc:Zero-or-many ;
	oslc:propertyDefinition foaf:knows ;
	oslc:range foaf:Person ;
	oslc:valueShape <associate> .
\end{lstlisting}

Listing~\ref{associate} contains the resource shape for associates.
\begin{lstlisting}[caption={Resource shape for associates}, label=associate]
# http://example.org/shapes/associate
@prefix foaf: <http://xmlns.com/foaf/0.1/> .
@prefix oslc: <http://open-services.net/ns/core#> .
@prefix xsd: <http://www.w3.org/2001/XMLSchema#>.
@base <http://example.org/shapes/> .

<associate> a oslc:ResourceShape ;
	oslc:describes foaf:Person ;
	oslc:property 
		<associate#name> , 
		<associate#isKnownBy> .

<associate#name> a oslc:Property ;
	oslc:name "name" ;
	oslc:occurs oslc:Exactly-one ;
	oslc:propertyDefinition foaf:name ;
	oslc:valueType xsd:string .

<associate#isKnownBy> a oslc:Property ;
	oslc:name "isKnownBy" ;
	oslc:occurs oslc:Exactly-one ;
	oslc:propertyDefinition foaf:knows ;
	oslc:isInverseProperty true ;
	oslc:range foaf:Person ;
	oslc:valueShape <contact> .
\end{lstlisting}

This example illustrates recursive shapes since the contact shape refers to the associate shape and the associate
shape refers to the contact shape. This apparent circularity would cause difficulty if contact and associate were
each described as a constraint. However, using the preceding formalism, the composite shape consisting of both the contact and
associate resource shapes is given a well-defined meaning.

In Resource Shape 2.0, neighbour functions are limited to forward and backward path expressions.
In fact, backward path expressions were missing from the original OSLC Resource Shape language and are a proposed
extension in Resource Shape 2.0. The proposed syntax for backward path expressions uses the optional property
{\tt oslc:isInverseProperty} but this design could be improved to provide better compatibility with downlevel clients,
i.e. a downlevel client might silently ignore this new property and produce incorrect results.

The following SPARQL query extracts the neighbour graph arcs from a set of resource shapes.
 Each binding of {\tt (?a ?direction ?p ?b)} corresponds to an arc $(a, q, b)$ where $q=forward(p)$ or $q=backward(p)$.
\begin{lstlisting}[caption={Query for neighbour graph arcs}, label=arcs]
prefix oslc: <http://open-services.net/ns/core#>

select distinct ?a ?direction ?p ?b
where {
	?a a oslc:ResourceShape ;
		oslc:property ?prop .
	?prop a oslc:Property ;
		oslc:propertyDefinition ?p ;
		oslc:valueShape ?b .
	optional {?prop oslc:isInverseProperty ?inverse}
	bind (if(bound(?inverse) && ?inverse, 
		'backward', 'forward') as ?direction)
}
\end{lstlisting}

The result of running this query on the PIM resource shapes is given in Table~\ref{result}.
\begin{table}[h]
\begin{center}
\begin{tt}
\begin{tabular}{|c|c|c|c|}
\hline
a			& direction	& p			& b \\
\hline
<contact>		& "forward"	& foaf:knows	& <associate> \\
<associate>	& "backward"	& foaf:knows	& <contact> \\
\hline
\end{tabular}
\end{tt}
\end{center}
\caption{Result of query on PIM shape}
\label{result}
\end{table}

The query correctly extracts the neighbour graph $pim\_ng$ of the PIM shape.

\subsubsection{Example: Polentoni}
Listing~\ref{polentoni-shape} contains the resource shape for Polentoni.
\begin{lstlisting}[caption={Resource shape for Polentoni}, label=polentoni-shape]
# http://example.org/shapes/polentoni
@prefix ex: <http://example.org/polentoni#> .
@prefix oslc: <http://open-services.net/ns/core#> .
@base <http://example.org/shapes/> .

<polentoni> a oslc:ResourceShape ;
	oslc:property 
		<polentoni#livesIn> , 
		<polentoni#knows> .

<polentoni#livesIn> a oslc:Property ;
	oslc:name "livesIn" ;
	oslc:occurs oslc:Exactly-one ;
	oslc:propertyDefinition ex:livesIn ;
	oslc:allowedValue ex:NorthernItaly .

<polentoni#knows> a oslc:Property ;
	oslc:name "knows" ;
	oslc:occurs oslc:Zero-or-many ;
	oslc:propertyDefinition ex:knows ;
	oslc:valueShape <polentoni> .
\end{lstlisting}

This resource shape is clearly recursive since it refers to itself.
However, as shown above, this recursion has a well-defined meaning and causes no difficulties.

\subsection{Relation to ShEx}

One major difference between Resource Shape 2.0 and ShEx is that ShEx allows the definition of
much richer node constraints using regular expressions, disjunction, and other operations.
ShEx also allows recursion via reference to named shapes, e.g. {\tt @<UserShape>},
which is referred to as the {\tt ValueReference} feature.

Most features of Resource Shape 2.0 can be expressed in ShEx. The intersection
of Resource Shape 2.0 and ShEx certainly includes the recursive aspects of Resource Shape 2.0 as
expressed by {\tt oslc:valueShape}.
Therefore the preceding formalism applies to ShEx provided that {\tt ValueReference} is used in a way that
maps directly to Resource Shape 2.0.

However, ShEx allows a more permissive use {\tt ValueReference}. 
For example, a {\tt ValueReference} may appear inside {\tt GroupRule} with cardinalities.
It is not clear that this usage can be expressed using suitably defined neighbour functions.

\subsection{Relation to SHACL}

The W3C Data Shapes Working Groups is currently developing the SHACL specification.
At the time of writing, there are three competing proposals.
One proposal, influenced by SPIN, appears to treat recursion similarly to Resource Shape 2.0.
A second proposal is a further development of ShEx.
The third proposal, SHACL-SPARQL, takes a different approach by avoiding recursion entirely.

The ability to name and refer to shapes allows for more intuitive descriptions of the constraints on graphs.
The absence of recursion in SHACL-SPARQL therefore detracts from its usefulness.
However, since the formalism presented here gives a clear and unobjectionable meaning to a
limited form of recursion, this capability could be added to SHACL-SPARQL.

\section{Conclusion}
\label{sec-conclusion}

The formalism presented here gives a precise meaning to recursive shapes as defined in Resource Shape 2.0.

This formalism is applicable to a subset of ShEx in which recursion is suitably limited.
More analysis is required in order to determine if the unlimited form of recursion allowed in ShEx and its SHACL follow-on
can be described using suitable neighbour functions, or if some new concept is required.

Finally, the limited form of recursion presented here could be added to the SHACL-SPARQL proposal to enhance its
expressiveness.

\cbstart
\section*{Acknowledgements}

Peter Patel-Schneider carefully reviewed an early draft of this article and provided valuable comments and criticisms.
Holger Knublauch suggested the use of the Polentoni example as a better illustration of recursion.

\cbend

\bibliography{shape-recursion}

\end{document}